\documentclass[twocolumn,aps,showpacs,superscriptaddress,floatfix,
               preprintnumbers,amsmath,amssymb,nofootinbib]{revtex4}

\usepackage{graphicx}
\usepackage{dcolumn}
\usepackage{bm}

\newcommand{\be}{\begin{equation}}      
\newcommand{\ee}{\end{equation}}        
\newcommand{\bear}{\begin{eqnarray}}    
\newcommand{\eear}{\end{eqnarray}}      
\newcommand{\beqstar}{\begin{eqnarray*}}        
\newcommand{\eeqstar}{\end{eqnarray*}}

\newcommand{\met}{\not \!\! E_T}
\newcommand{\deltat}{\hat{\delta}}
\newcommand{\deltab}{\bar{\delta}}

\begin{document}

\preprint{EFI-02-80, UFIFT-HEP-02-7, CERN-TH/2002-108, BUHEP-02-21}

\title{Bosonic Supersymmetry? Getting Fooled at the LHC}

\author{Hsin-Chia Cheng}
\affiliation{Enrico Fermi Institute, 
The University of Chicago, Chicago, IL 60637, USA}
\author{Konstantin T. Matchev}
\affiliation{Department of Physics, University of Florida,
Gainesville, FL 32611, USA}
\affiliation{TH Division, CERN, Geneva 23, CH--1211, Switzerland}
\author{Martin Schmaltz}
\affiliation{Department of Physics, Boston University,
Boston, MA 02215, USA}

\date{May 28, 2002}

\begin{abstract}
We define a minimal model with Universal Extra Dimensions, and begin
to study its phenomenology. The collider signals of the first KK level
are surprisingly similar to those of a supersymmetric model with a nearly
degenerate superpartner spectrum. The lightest KK particle (LKP)
is neutral and stable because of KK-parity. KK excitations cascade
decay to the LKP yielding missing energy signatures with relatively 
soft jets and leptons. Level 2 KK modes may also be probed via their
KK number violating decays to Standard Model particles. In either case 
we provide initial estimates for the discovery potential of the Tevatron
and the LHC.

\end{abstract}

\maketitle

\section{\label{sec:intro}Introduction}

The new ideas of extra dimensions and localized gravity
have recently attracted a lot of interest. They not only
offer exciting new avenues for theoretical exploration
but also predict
signals which can soon be tested at the upcoming collider 
experiments at the Fermilab Tevatron and the CERN 
Large Hadron Collider (LHC).

The focus of this paper is on Universal Extra Dimensions
(UEDs)~\cite{Appelquist:2001nn},
a model in which all Standard Model fields
propagate in extra dimensions of size $R^{-1}\sim$ TeV. Although
there are many theoretical reasons for studying UEDs
(electroweak symmetry breaking~\cite{Arkani-Hamed:2000hv}, 
proton decay~\cite{Appelquist:2001mj},
the number of generations~\cite{Dobrescu:2001ae},
neutrino masses~\cite{Appelquist:2002ft}, etc.),
we are primarily motivated by their collider phenomenology.
Experimental bounds allow Kaluza-Klein (KK) modes in UEDs to
be as light as a few hundred 
GeV \cite{Appelquist:2001nn,Agashe:2001xt,Appelquist:2001jz}.
The production cross section
at the LHC for KK excitations of quarks and gluons
weighing only a few hundred GeV is enormous. However,
as we discuss in this paper, their subsequent detection is
non-trivial because they decay nearly invisibly.
The phenomenology of UEDs shows interesting parallels to
supersymmetry. Every Standard Model field has KK partners. The
lowest level KK partners carry a conserved quantum number,
KK parity, which guarantees that the lightest KK particle (LKP) is stable.
Heavier KK modes cascade decay to the LKP by emitting soft
Standard Model particles. The LKP escapes detection, resulting
in missing energy signals. 

In the following section we define Minimal Universal Extra
Dimensions (MUEDs). The model is defined in five
dimensions with one dimension
compactified on an $S_1/Z_2$ orbifold. All fields propagate in
the bulk and have KK modes with masses approximately equal to
the compactification
scale. The Lagrangian of the model includes interactions which are localized
at the boundaries of the orbifold. These boundary
terms lead to mass splittings between KK modes and
affect their decays. 
In Sections III (and IV) we discuss the phenomenology of the first 
(and second) level KK states. We identify possible decay modes
and branching ratios, and we estimate the discovery reach at the
Tevatron and the LHC. Section V contains our conclusions and
speculations about the cosmology of UEDs.

\section{Minimal Universal Extra Dimensions}
\label{sec:model}

The simplest UED scenario has all of the Standard Model fields
(no supersymmetry) propagating in a single extra dimension.
In 4+1 dimensions, the fermions 
[$Q_i,\, u_i, \, d_i,\, L_i,\, e_i, \, i=1,2,3$, where upper (lower) case 
letters represent $SU(2)$ doublets (singlets)] are four-component and contain
both chiralities when reduced to 3+1 dimensions. To produce a chiral
4d spectrum, we compactify the extra dimension on an $S_1/Z_2$ orbifold.
Fields which are odd under the $Z_2$ orbifold symmetry
do not have zero modes, hence the unwanted fields (zero modes of fermions with
the wrong chiralities and the 5th component of the gauge fields)
can be projected out. The remaining zero modes are just the Standard Model
particles in 3+1 dimensions.

The full Lagrangian of the theory comprises both
bulk and boundary interactions.
Gauge and Yukawa couplings and the Higgs potential are contained in the
bulk Lagrangian in one-to-one correspondence with
the couplings of the Standard Model. The boundary
Lagrangian interactions are localized at the orbifold
fixed points and do not respect five dimensional Lorentz invariance.

Ignoring the localized terms for the moment, the mass of the $n$-th KK
mode is
\begin{equation}
\label{tree-spectrum}
m_n^2= \frac{n^2}{R^2} + m_0^2,
\end{equation}
where $R$ is the radius of the compact dimension, and $m_0$ is the zero
mode mass. The spectrum at each KK level is highly degenerate except
for particles with large zero mode masses ($t,\,  W,\, Z,\, h$).
The bulk interactions preserve the 5th dimensional momentum
(KK number). The corresponding coupling constants among KK modes 
are simply equal to
the SM couplings (up to normalization factors such as $\sqrt{2}$).
The Feynman rules for the KK modes can easily be derived 
(e.g., see Ref.~\cite{Macesanu:2002db,Dicus:2000hm}).

In contrast, the coefficients of the boundary terms are not
fixed by Standard Model couplings and correspond to new
free parameters. In fact, they are renormalized by the bulk interactions
and hence are scale dependent \cite{Georgi:2001ks,Cheng:2002iz}.
One might worry
that this implies that all predictive power is lost.
However, since the wave functions of Standard Model fields and KK modes are
spread out over the extra dimension and the new couplings only exist on the
boundaries, their effects are volume suppressed. We can get an
estimate for the size of these volume suppressed corrections with naive
dimensional analysis by assuming strong coupling at the cut-off.
The result is that the mass shifts to KK modes from boundary terms
are numerically equal to corrections from loops
$\delta m_n^2/m_n^2 \sim g^2/16 \pi^2$.

We will assume that the boundary terms are symmetric under the exchange
of the two orbifold fixed points, which preserves the KK parity 
discussed below. Most relevant to the phenomenology are localized kinetic
terms for the SM fields, such as
\begin{equation}
\label{newops}
\frac{\delta(x_5)+\delta(x_5-\pi R)}{\Lambda} 
\left[
G_4 (F_{\mu\nu})^2 
+ F_4 \overline \Psi i \slash\!\!\!\!D \Psi + 
F_5 \overline \Psi \gamma_5 \partial_5 \Psi
\right], 
\end{equation}
where the dimensionless coefficients $G_4$ and $F_i$ are arbitrary and
not universal for the different Standard Model
fields. These terms are important phenomenologically for several 
reasons: ({\it i}) they split the near-degeneracy of KK modes at each level,
({\it ii})
they break KK number conservation down to a KK parity
under which modes with odd KK numbers are charged,
({\it iii}) they introduce possible new flavor violation.

Since collider signatures depend strongly on the values of the
boundary couplings it is necessary to be definite and specify them.
A reasonable ansatz is to take flavor-universal boundary terms.
Non-universalities would give rise to FCNCs as in supersymmetry
with flavor violating scalar masses. This still leaves a large number of
free parameters.
For definiteness, and also because we find the
resulting phenomenology especially interesting,
we make the assumption that all boundary 
terms are negligible at some scale $\Lambda > R^{-1}$.
This defines our model.

Note that this is completely analogous to the case of the 
Minimal Supersymmetric Standard Model (MSSM) where one has to choose
a set of soft supersymmetry breaking couplings at some high scale,
before studying the phenomenology. Different ansaetze for the parameters
can be justified by different theoretical prejudices but ultimately
one should use experimental data to constrain them. In a sense,
our choice of boundary couplings may be viewed as analogous to the
simplest minimal supergravity boundary condition --
universal scalar and gaugino masses.
Thus the model of MUEDs is extremely predictive and has only 
three free parameters: 
\begin{equation}
\{R,\Lambda,m_h\}\ ,
\label{eq:parameters}
\end{equation}
where $m_h$ is the mass of the Standard Model Higgs boson.

The low energy KK spectrum of MUEDs depends on the boundary terms
at low scales which are determined from the
high energy parameters through the renormalization group. Since
the corrections are small we use the one-loop leading log
approximations. In addition to the boundary terms we also take
into account the non-local radiative corrections to KK masses. All
these were computed at one-loop in \cite{Cheng:2002iz}.

\begin{figure}[tb]
\includegraphics[width=.48\textwidth]{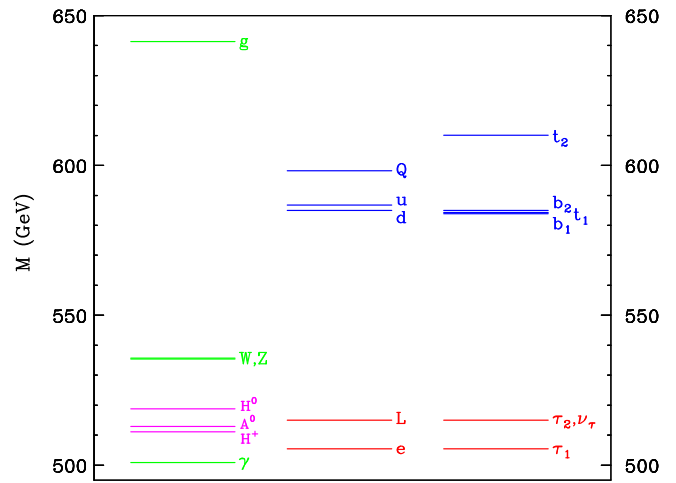}
\caption{\label{fig:spectrum} {\it One-loop corrected mass spectrum of
the first KK level in MUEDs for $R^{-1}=500$ GeV, $\Lambda R = 20$
and $m_h=120$ GeV.}}
\end{figure}
\begin{figure}[tb]
\includegraphics[height=2.5in]{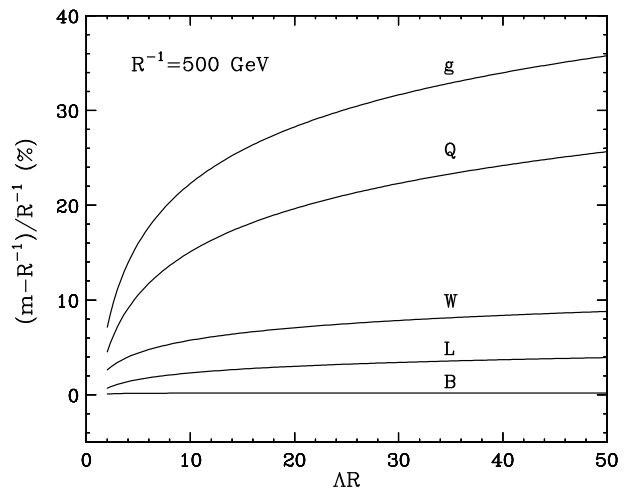}%
\caption[]{\it Radiative corrections (in \%)
to the spectrum of the first KK level for $R^{-1}=500$ GeV,
versus $\Lambda R$.}
\label{fig:corrections}
\end{figure}
A typical spectrum for the first level KK modes is shown in
Fig.~\ref{fig:spectrum}.
Fig.~\ref{fig:corrections} shows the dependence of 
the splittings between first level KK modes on the cutoff scale $\Lambda$.
Typically, the corrections for KK modes with strong interactions are 
$> 10\%$ while those for states with only electroweak interactions are a few
percent. We find that the corrections to the masses are such that 
$m_{g_n}>m_{Q_n}>m_{q_n}>m_{W_n}\sim m_{Z_n}>m_{L_n}>m_{\ell_n}>m_{\gamma_n}$.
The lightest KK particle $\gamma_1$, is a mixture of the first KK 
mode $B_1$ of the $U(1)_Y$
gauge boson $B$ and the first KK mode $W^0_1$ of the $SU(2)_W$ $W^3$ 
gauge boson.
(The possibility of the first level KK graviton being the LKP is irrelevant
 for collider phenomenology, since the decay lifetime of $\gamma_1$
 to $G_1$ would be of cosmological scales.)
We will usually denote this state by $\gamma_1$. However, note that
the corresponding ``Weinberg'' angle $\theta_1$ is much smaller than
the Weinberg angle $\theta_W$ of the Standard Model \cite{Cheng:2002iz},
so that the $\gamma_1$ LKP is mostly $B_1$ and $Z_1$ is mostly $W^0_1$.
The mass splittings among the level 1 KK modes are large enough 
for the prompt decay of a heavier level 1 KK mode to a lighter 
level 1 KK mode. But since the spectrum
is still quite degenerate, the ordinary SM particles emitted from
these decays will be soft, posing a challenge for collider searches.
 
The terms localized at the orbifold fixed points also violate the KK number 
by even units. However, assuming that no explicit KK-parity violating 
effects are put in by hand, KK parity remains an exact symmetry.
The boundary terms allow higher ($n>1$) KK modes to decay to lower KK modes,
and even level states can be singly produced (with smaller cross sections
because the boundary couplings are volume suppressed). Thus
KK number violating boundary terms are important for higher KK mode searches
as we will discuss in Section~\ref{sec:second}.

\section{First KK level}
\label{sec:first}

\begin{figure}[tb]
\includegraphics[height=2.5in]{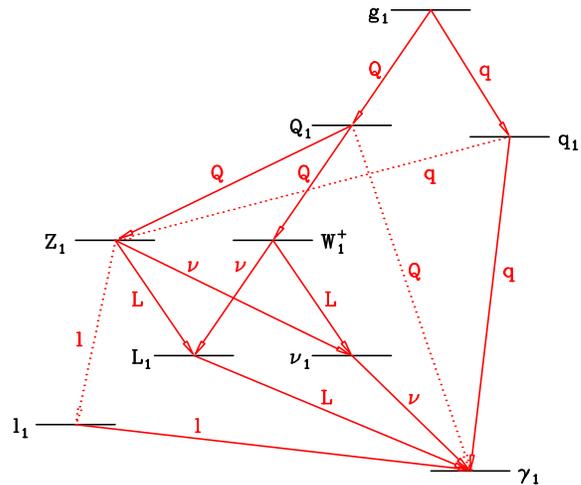}
\caption[]{\em Qualitative sketch of the level 1
KK spectroscopy depicting the dominant (solid) and rare (dotted) 
transitions and the resulting decay product.
\label{fig:transitions}}
\end{figure}

Once the radiative corrections are included, the KK mass degeneracy at 
each level is lifted and the KK modes decay promptly. The collider
phenomenology of the first KK level is therefore very similar to
a supersymmetric scenario in which the superpartners are relatively
close in mass - all squeezed within a mass window of 100-200 GeV
(depending on the exact value of $R$). Each level 1 KK particle has
an exact analogue in supersymmetry: $B_1\leftrightarrow$ bino,
$g_1\leftrightarrow$ gluino, $Q_1(q_1)\leftrightarrow$ left-handed
(right-handed) squark, etc. The decay cascades of the level 1 KK modes
will terminate in the $\gamma_1$ LKP (Fig.~\ref{fig:transitions}).
Just like the neutralino LSP is stable in $R$-parity conserving
supersymmetry, the $\gamma_1$ LKP in MUEDs is stable due to KK parity 
conservation and its production at colliders results in generic
missing energy signals. 

It is known that supersymmetry with a stable neutralino LSP
is difficult to discover at hadron colliders if the superpartner
spectrum is degenerate. Hence the discovery of level 1 
KK modes in MUEDs at first sight appears problematic as well -- the decay
products resulting from transitions between level 1 KK states may be too
soft for reliable experimental observation at hadron colliders.
This issue is the subject of this Section.

Before we address the possible level 1 discovery channels in some detail,
we need to determine the allowed decays at level 1 and estimate
their branching fractions. For any given set of input parameters
(\ref{eq:parameters}) the mass spectrum and couplings of the KK modes 
in MUEDs are exactly calculable~\cite{Cheng:2002iz}. Hence one obtains
very robust predictions for the main branching ratios of interest for
phenomenology.

{\em KK gluon.}--- The heaviest KK particle at level 1 is the KK gluon $g_1$.
Its two-body decays to KK quarks $Q_1$ and $q_1$ are always open
and have similar branching fractions:
$B(g_1\to Q_1 Q_0) \simeq B(g_1\to q_1 q_0) \simeq 0.5$.

{\em KK quarks.}--- The case of $SU(2)$-singlet quarks ($q_1$)
is very simple -- they can only decay to the hypercharge gauge boson
$B_1$, hence their branchings to $Z_1$ are suppressed by
the level 1 Weinberg angle $\theta_1\ll \theta_W$:
$B(q_1\to Z_1 q_0) \simeq \sin^2\theta_1 \sim 10^{-2}-10^{-3}$
while $B(q_1\to \gamma_1 q_0) \simeq \cos^2\theta_1 \sim 1$.
Thus $q_1$ production yields jets plus missing energy,
the exception being $t_1\to W^+_1 b_0$ and $t_1\to H^+_1 b_0$
(the latter will be in fact the dominant source of $H^+_1$ production
at hadron colliders). 

$SU(2)$-doublet quarks ($Q_1$) can decay to
$W^\pm_1$, $Z_1$ or $\gamma_1$. In the limit $\sin\theta_1\ll1$
$SU(2)_W$-symmetry implies
\begin{equation}
B(Q_1\to W^\pm_1 Q'_0) \simeq 2 B(Q_1\to Z_1 Q_0)
\label{eqn:Q1Ws}
\end{equation}
and furthermore for massless $Q_0$ we have
\begin{equation}
\frac{B(Q_1\to Z_1 Q_0)}{B(Q_1\to \gamma_1 Q_0)} \simeq
\frac{g_2^2\,T_{3Q}^2\, (m_{Q_1}^2-m_{Z_1}^2)}
     {g_1^2\,Y_{ Q}^2\, (m_{Q_1}^2-m_{\gamma_1}^2)} \ ,
\label{eqn:Q1ratio}
\end{equation}
where $g_2$ ($g_1$) is the $SU(2)_W$ ($U(1)_Y$) gauge coupling,
and $T_3$ and $Y$ stand for weak isospin and hypercharge,
correspondingly. We see that the $Q_1$ decays to $SU(2)$
gauge bosons, although suppressed by phase space, are numerically 
enhanced by the ratio of the couplings and quantum numbers.
With typical values for the mass corrections from Fig.~\ref{fig:corrections},
eqs.~(\ref{eqn:Q1Ws}) and (\ref{eqn:Q1ratio}) yield
$B(Q_1\to W^\pm_1 Q'_0)\sim 65\%$, $B(Q_1\to Z_1 Q_0)\sim 33\%$ and
$B(Q_1\to \gamma_1 Q_0)\sim  2\%$. 

{\em KK $W$- and $Z$-bosons.}--- With their hadronic decays closed,
$W^\pm_1$ and $Z_1$ decay democratically to all lepton flavors:
$B(W^\pm_1\to \nu_1      L^\pm_0) = B(W^\pm_1\to L^\pm_1\nu_0) = \frac{1}{6}$
and
$B(Z_1\to \nu_1\bar{\nu}_0) = B(Z_1\to L^\pm_1 L^\mp_0) \simeq \frac{1}{6}$
for each generation. $Z_1\to\ell^\pm_1\ell^\mp_0$ decays are suppressed by
$\sin^2\theta_1$. 

{\em KK leptons.}--- The level 1 KK modes of the charged leptons as well as the
neutrinos decay directly to $\gamma_1$. As a result $W^\pm_1$ and $Z_1$
always effectively decay as $W^\pm_1\to \gamma_1 L^\pm_0 \nu_0$ and
$Z_1\to \gamma_1 L^\pm_0 L^\mp_0$ or $Z_1\to \gamma_1 \nu_0 \bar{\nu}_0$,
with relatively large $e$ and $\mu$ yields.

{\em KK Higgs bosons.}--- 
Their decays depend on their masses.
They can decay into the KK $W$, $Z$ bosons or KK $t$, $b$ quarks
if they are heavier and the phase space is open. On the other hand,
if they are lighter than $W_1$, $Z_1$, $t_1$, $b_1$
(as in the example of Fig.~\ref{fig:spectrum}), their tree-level two-body 
decays will be suppressed. Then they will decay to $\gamma_1$ and the 
corresponding virtual zero-level Higgs boson, or to $\gamma_1 \gamma_0$ 
through a loop.

We are now in shape to discuss the optimum strategy for MUEDs
KK searches at hadron colliders. Level 1 KK states necessarily
have to be pair produced, due to KK parity conservation. The
approximate mass degeneracy at each level ensures that strong
production dominates, with all three
subprocesses (quark-quark, quark-gluon and gluon-gluon)
having comparable rates~\cite{Rizzo:2001sd,Macesanu:2002db}.

For an estimate of the reach at the Tevatron or the LHC, we need
to discuss the final state signatures and the related backgrounds.
The signature with the largest overall rate is $\met+N\ge2$ jets, which is
similar to the traditional squark and gluino searches~\cite{squarks}.
It arises from inclusive (direct or indirect) $q_1 q_1$ production.
Roughly one quarter of the {\em total} strong production cross-section 
$\sigma^{had}_{tot}$ materializes in $q_1 q_1$ events.
However, in spite of the large missing mass in these events, the
{\em measured} missing energy is rather small, since it is correlated
with the energy of the relatively soft recoiling jets.
As a conservative rough guide for the discovery reach we can use
existing studies of the analogous supersymmetric case.
One might expect that Run II can probe $R^{-1}\sim300$ GeV~\cite{Abel:2000vs}
while the LHC reach for $R^{-1}$ is no 
larger than 1.2 TeV~\cite{Bityukov:1999am}.
While the jetty signatures can be potentially used for discovery,
further studies in an MUEDs context are needed. 
Here we prefer to discuss the much cleaner multilepton final states
arising from diboson ($W^\pm_1$ or $Z_1$) production.

Consider inclusive $Q_1Q_1$ production, whose cross-section
also roughly equals $\frac{1}{4}\sigma^{had}_{tot}$. The 
subsequent decays of $Q_1$'s yield $W^\pm_1W^\pm_1$, 
$W^\pm_1 Z_1$ and $Z_1Z_1$ pairs in proportion $4:4:1$.
The $W^\pm_1$ and $Z_1$ decays in turn provide multilepton
final states with up to 4 leptons plus missing energy,
all of which may offer the possibility of a discovery.
In the following we concentrate on the gold-plated $4\ell\met$ 
signature.

\begin{figure}[t]
\includegraphics[height=2.5in]{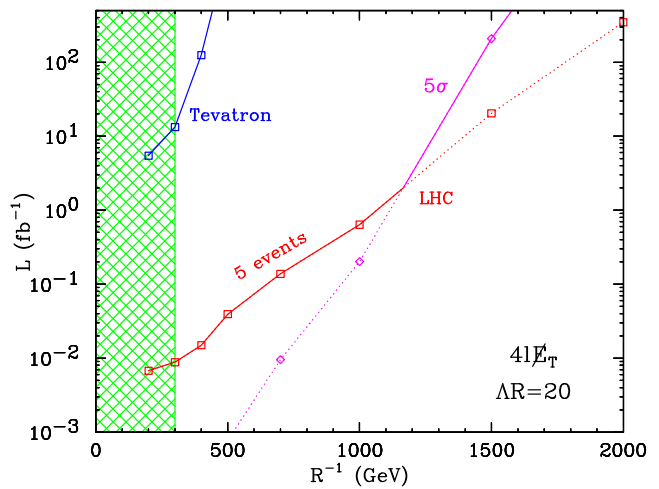}%
\caption[]{\it Discovery reach for MUEDs at
the Tevatron (blue) and the LHC (red) in the $4\ell\met$ channel. 
We require a $5\sigma$ excess or the observation of 5 signal events,
and show the required total integrated
luminosity per experiment (in ${\rm fb}^{-1}$) as a function
of $R^{-1}$, for $\Lambda R=20$. 
(In either case we do not combine the two experiments).}
\label{fig:reach}
\end{figure}

We shall conservatively ignore additional signal contributions from
direct diboson production and $Q_1W^\pm_1$ or $Q_1Z_1$ processes. 
For the Tevatron we use the single lepton triggers $p_T(\ell)>20$ GeV
and $|\eta(e)|<2.0$, $|\eta(\mu)|<1.5$; or the missing energy trigger
$\met>40$ GeV. Because the channel is very clean, we use
relatively soft off-line cuts, $p_T(\ell)>\{15,10,10,5\}$ GeV,
$|\eta(\ell)|<2.5$ and $\met>30$ GeV. The remaining physics
background comes from 
$ZZ\to \ell^\pm\ell^\mp\tau^+\tau^-\to 4\ell\met$
where $Z$ stands for a real or virtual $Z$ or $\gamma$~\cite{Matchev:1999yn},
and can be reduced by invariant mass cuts
for any pair of opposite sign, same flavor leptons: $|m_{\ell\ell}-M_Z|>10$ GeV
and $m_{\ell\ell}>10$ GeV. As a result, the expected background is less
than 1 event in all of Run II and we require 5 signal events for discovery.
The reach is shown in Fig.~\ref{fig:reach}.
We see that Run IIb of the Tevatron will
go slightly beyond the current indirect bounds
($R^{-1}>300$ GeV) from precision data \cite{Appelquist:2001nn}.

For the LHC we use $p_T(\ell)>\{35,20,15,10\}$ GeV with
$|\eta(\ell)|<2.5$, which is enough for the single lepton trigger.
In addition, we require $\met > 50$ GeV and the same dilepton 
invariant mass cut. There are now several relevant background
sources, including multiple gauge boson and/or top quark
production~\cite{Barger:1991jn}, fakes, leptons from $b$-jets etc.
We conservatively assume a background level of 50 events after cuts
per 100 ${\rm fb}^{-1}$ (1 year of running at high luminosity).
Our LHC reach estimate is presented in Fig.~\ref{fig:reach}.
Without combining experiments, we plot the total integrated luminosity
$L$ required for either an observation of 5 signal events or a $5\sigma$
excess over the background. The reach, shown as a solid line,
is defined as the larger of the two and extends to $R^{-1}\sim 1.5$ TeV. 

Other leptonic channels such as two or three leptons with $\met$ may
also be considered. They have more backgrounds but take
advantage of the larger branching fraction for $Q_1\to W^\pm_1Q'_0$
and offer higher statistics, which may prove useful especially 
for the case of the Tevatron.

In conclusion, note that at a hadron collider all signals from
level 1 KK states look very much like supersymmetry --
all SM particles have ``partners'' with similar couplings,
and identifying the extra-dimensional nature of the new physics
becomes rather challenging. Nevertheless, there are three features
which distinguish the MUEDs scenario from ordinary
supersymmetry. First, the spins are different, but this ``bosonic''
nature of the newly discovered ``supersymmetry'' will most likely
escape detection at a hadron collider. Second, the analogy with the MSSM 
is incomplete, as MUEDs do not have analogues of the
``heavy'' Higgs bosons of the MSSM. To be more precise, the level 1
KK modes of the Higgs have exactly the same gauge quantum numbers as the
MSSM Higgs bosons $H^0, A^0, H^{\pm}$. But since they carry KK parity,
their behavior is similar to that of higgsinos instead.
Now recall that there are regions of the MSSM parameter space
where the LHC can only discover the SM-like Higgs boson, and
misses the other three Higgs states of the MSSM.
MUEDs could  easily be confused with this scenario.
This leaves us with the single smoking gun signature for MUEDs --
the presence of higher level KK modes. 

\section{Second KK level}
\label{sec:second}

Through KK number preserving interactions a level 2 KK state can decay
to two level 1 KK modes, or to another level 2 KK state and a
SM particle. For example, the level 2 fermion
decay widths (for massless $f_0$ and in leading order of $\hat{\delta} m$)
are easily computed at tree level using the Feynman rules for KK modes.
We find 
\begin{equation}
\Gamma(f_2\to V_2\, f_0) \approx \frac{3c^2 g^2 m_{f_2}}{8\pi}
\left( \frac{\deltat m_{f_2}}{m_2} -  \frac{\deltat m_{V_2}}{m_2}
\right)^2, \label{f2toV2}
\end{equation}
\vspace{-0.3cm}
\begin{equation}
\Gamma(f_2\rightarrow V_1\, f_1) \approx  
\frac{11c^2 g^2 m_{f_2}}{16\sqrt{2}\pi}
\left( \frac{\deltat m_{f_2}}{ m_2} 
     - \frac{\deltat m_{V_1}}{2m_1} 
     - \frac{\deltat m_{f_1}}{2m_1}
\right)^{\frac{3}{2}}, \label{f2toV1}
\end{equation}
where $\deltat m$ represents the total mass correction
and $c$ is a Clebsch factor. For a level 2 gauge boson
\begin{equation}
\Gamma(V_2\to f_2\, f_0)  \approx  \frac{c^2 g^2 m_{V_2}}{4\pi}
\left( \frac{\deltat m_{V_2}}{m_2} 
     - \frac{\deltat m_{f_2}}{m_2}
\right)^2, \label{V2tof2}
\end{equation}
\vspace{-0.3cm}
\begin{equation}
\Gamma(V_2\to f_1\, f'_1)  \approx  \frac{c^2 g^2 m_{V_2}}{6\sqrt{2}\pi}
\left( \frac{\deltat m_{V_2}}{m_2} 
     - \frac{\deltat m_{f_1}}{2m_1}
     - \frac{\deltat m_{f'_1}}{2m_1}
\right)^{\frac{3}{2}}, \label{V2tof1}
\end{equation}
counting both KK chiralities in the last case.
All of the decays (\ref{f2toV2}-\ref{V2tof1}) are phase space suppressed,
once again leaving rather little visible energy
deposited in the detector. 

A level 2 KK gauge boson, however, can also decay
directly to two SM particles via KK number violating
interactions~\cite{Cheng:2002iz}. The width is
\begin{equation}
\Gamma(V_2\to f_0\, f_0)  \approx  \frac{c^2 g^2 m_{V_2}}{12\pi}
\left( \frac{\deltab m_{V_2}}{m_2} 
    -  \frac{\deltab m_{f_2}}{m_2}
\right)^2,
\label{V2to00}
\end{equation}
where $\deltab m$ only contains the mass corrections due to
the boundary terms (though typically $\deltat m \simeq \deltab m$.)
These decays are {\em not} phase space suppressed, and deposit
a lot of energy, hence they offer the best opportunity for a level 2 discovery.

Level 2 KK gauge bosons can be pair-produced through KK number preserving
interactions, or singly produced through their suppressed KK number
violating couplings to SM quarks and leptons. We first concentrate on
the $g_2$ signal. Using eqs.~(\ref{V2tof2}-\ref{V2to00}), we find
$B(g_2\to Q_0Q_0,q_0q_0)\simeq 0.1$. The production of $g_2$ in
association with another level 2 colored particle then yields a
unique $\met+N>2$ jet signature, where the invariant mass of the
two leading jets reconstructs to $m_{g_2}$. In the absence of a
problematic physics background, we require 10 events before cuts
and efficiencies for discovery, leaving us with a reach for $R^{-1}$
of just below 1 TeV. $W^\pm_2$, $Z_2$ and $\gamma_2$ searches in
their hadronic modes will be very similar. Branching fractions to
leptonic decay modes are very small and do not permit a significant reach.
Notice that $\gamma_2$ has no KK preserving decay modes left open,
hence $\sum_f B(\gamma_2\to f_0f_0)\simeq 1$.

The usual $W'/Z'$ and coloron searches are sensitive to singly produced level
2 KK gauge bosons. 
However, the reach is inferior due to the smallness of
the KK number violating couplings, which are only a fraction of the
SM gauge couplings -- typically 10-20\% for quarks and only a few
percent for leptons.

\section{ conclusions} 
\label{sec:conclusions}

Universal extra dimensions with compactification radius near the
TeV scale promise exciting phenomenology for future colliders.
All Standard Model particles have KK partners which can be
produced with enormous cross sections at the LHC. As we showed
in this paper, the detection of KK particles at the LHC is non-trivial
as they decay to very soft Standard Model particles which
are difficult (but not impossible) to see above background.
Clearly, more realistic simulations of the phenomenology of
MUEDs are necessary and studies for different  values of the
boundary couplings would be of interest as well.

A lepton collider running at the center of mass energy of the second
level photon or $Z$ is ideal for measuring the small
mass splittings between states and determining spins. However,
the required center of mass energy ($\sim 2\ R^{-1}$) may be
too high for the next generation linear collider. 

Finally, we note that similarly to the neutralino LSP in supersymmetry,
the $\gamma_1$ LKP of MUEDs is a great cold dark matter candidate,
whose annihilation rate is {\em not} helicity suppressed.
A study of the resulting abundance and detection opportunities
is underway \cite{darkm}.

\begin{acknowledgments}
We would like to thank M.~Chertok, B.~Dobrescu
and B.~Schumm for useful discussions, and D.~E.~Kaplan for suggesting
the catchy title. We also thank the Aspen Center for Physics
for hospitality during the initial stage of this work.
H.-C. C. is supported by the Department of Energy grant DE-FG02-90ER-40560.  
M.S. is supported in part by the Department of Energy under grant
number DE-FG02-91ER-40676.
\end{acknowledgments}


\end{document}